\documentclass[journal=jacsat,manuscript=article]{achemso}
\setkeys{acs}{articletitle = true}
\setkeys{acs}{maxauthors = 0}
\usepackage[version=3]{mhchem} 

\usepackage{xcolor,colortbl}
\usepackage{color}
\usepackage{braket}
\usepackage{commath}
\usepackage{tablefootnote}
\usepackage{adjustbox}
\usepackage{chemfig}
\usepackage{subcaption}
\usepackage[skip=2pt]{caption} 
\usepackage{url}
\usepackage{hyperref}
\usepackage{soul}
\usepackage{multicol}
\usepackage{multirow}

\author{Rajendra P. Joshi}
\affiliation[{Department of Computer Science, Department of Physics,  Science of Advanced Materials, Central Michigan University, Mount Pleasant, MI, 48859, USA}]
{Department of Physics and Science of Advanced Materials Program, Central Michigan University, Mount Pleasant, MI, 48859, USA}
\alsoaffiliation[{Department of Computer Science,  Central Michigan University, Mount Pleasant, MI, 48859, USA}]
{Department of Computer Science,  Central Michigan University, Mount Pleasant, MI, 48859, USA}

\author{Jesse Eickholt}
\author{Liling Li}
\affiliation[{Department of Computer Science,  Central Michigan University, Mount Pleasant, MI, 48859, USA}]
{Department of Computer Science,  Central Michigan University, Mount Pleasant, MI, 48859, USA}

\author{Marco Fornari}
\author{Veronica Barone}
\email{v.barone@cmich.edu}
\author{Juan E. Peralta}
\affiliation[{Department of Physics,  Science of Advanced Materials Program, Central Michigan University, Mount Pleasant, MI, 48859, USA}]
{Department of Physics and Science of Advanced Materials Program, Central Michigan University, Mount Pleasant, MI, 48859, USA}

\title{Machine Learning the Voltage of Electrode Materials in Metal-ion Batteries}

\abbreviations{IR,NMR,UV}

\begin{document}

\begin{abstract}
Machine learning (ML) techniques have rapidly found applications in many domains of materials chemistry and physics where  large data sets are available. 
Aiming to accelerate the discovery of materials for battery applications, in this work, we develop a tool (\url{http://se.cmich.edu/batteries}) based on ML models to predict voltages of electrode materials for metal-ion batteries.
To this end, we use deep neural network, support vector machine, and kernel ridge regression as ML algorithms in combination with data taken from the Materials Project Database, as well as feature vectors from properties of chemical compounds and elemental properties of their constituents.
We show that our ML models have predictive capabilities for different reference test sets and, as an example, we utilize them to generate a voltage profile diagram and compare it to density functional theory calculations.
In addition, using our models, we propose nearly 5,000  candidate electrode materials for Na- and K-ion batteries. 
We also make available a web-accessible tool that, within a minute, can be used to estimate the voltage of any bulk electrode material for a number of metal-ions. These results show that ML is a promising alternative for computationally demanding calculations as a first screening tool of novel materials for battery applications.
\end{abstract}

\section{Keywords}
 Machine learning, batteries, intercalation electrodes, web-tool, voltage predictor, voltage profile diagram

\section{Introduction}

Lithium-ion batteries (LIBs) have  revolutionized the energy storage technology\cite{before-Li-ion, battery-future} and played a crucial role in transforming portable devices in terms of performance, lifetime, weight, and size. They have also opened unprecedented possibilities for new greener technologies, for instance in the automotive sector.\cite{ MAX, battery-future}
Despite currently being the dominating energy source for small energy scale devices, the transferability of traditional LIBs to higher energy scales remains a challenge mainly because of their relatively low energy density.\cite{LiNaK, Al-BC3, gridscale, grid}
Moreover, the future of large scale applications of LIBs is uncertain due to the scarcity of Li as a raw material, its increasing price, skyrocketing energy demand, and safety concerns.\cite{batterysafety, tarascon2011issues, tarascon2010lithium}
These issues call for more sustainable, cheaper, and better performing alternatives to the present technology.

Several new designs for metal-ion batteries have been proposed in the literature, including monovalent Na- and K-ion
batteries, and multivalent Mg-, Ca-, and Al-ion batteries.\cite{LiNaK, Al-BC3, MAX, Multivalent, LiMgAL, BeyondLi}
The development of these batteries has been limited mainly because of a lack of suitable electrode and electrolyte materials, stimulating
considerable work devoted to enhancing their robustness\cite{Rajput2018-electrolytes}.
For electrode materials, the number of
possible compounds that could intercalate metal ions such as Li, Na, K,
Mg, Ca, and Al is likely in the order of thousands;
however, the majority of these materials
remain unexplored as electrode components because of the experimental and 
computational difficulties to screen the large chemical and structural space with appropriate accuracy.\cite{chemicalspace, chemicalspace3, chemicalspace2}
In this context, data driven machine learning (ML) approaches provide ways to address this issue faster and with limited use of computational resources.\cite{ML1, ML2}

ML-based search for new electrode materials requires a sufficient amount of well curated   and verified data.\cite{roadmap}  
Recently, enabled by the significant improvement in computing architectures,
several databases based on density functional theory (DFT) calculations such as AFLOW\cite{aflow, AFLOW-api}, Materials
Project\cite{Materials_project, materials_API}, OQMD,\cite{OQMD1, OQMD2} and NOMAD\cite{NOMAD} 
were made readily available to the scientific community. Although DFT predictions are not the  gold standard for theoretical calculations in some contexts,\cite{SIE-J}  they do provide reasonable insights and can be used to guide experimental research.\cite{BN-doped-C}
In several cases, these electronic structure databases have been combined with ML approaches for predicting  specific properties of interest in  target materials\cite{ML-review, ML-layeredMater, ML-strategy, chemicalspace3,  ML-7,  ML-review-accepted, AFLOW-api}.
For example,
Seko \textit{et al.}\cite{Seko, seko1} used the kernel ridge regression (KRR) and support vector regression (SVR) to predict the
cohesive energy of binary and ternary compounds, thermal conductivity of binary inorganic compounds, and melting temperature of single- and binary-component solids. Similarly, Meredig \textit{et al.}\cite{Meredig} built a ML model to
predict the thermodynamic stability of any arbitrary  chemical composition and proposed nearly 4500 new stable chemical
compounds.  ML approaches have also been used 
to improve the quality of exchange-correlation functionals in DFT. \cite{bypassing-KS-ML, ML-qm, ML-functionals}
These approaches have also been applied for the prediction of the band gap, total energy, and to find
the potential candidate materials for photovoltaic cells, glass alloys, \textit{etc.} \cite{ward2016general, energyelapsolite, LR-DFT}
Logistic regression was used by Sendek \textit{et al.}\cite{MLelectrolytes} to
screen nearly 12,000 solids containing Li to propose new  materials for electrolytes in Li-ion batteries.
Similarly, online tools for ML predictions of electronic, thermal, and mechanical properties have been made available lately.\cite{AFLOW-api}

In this work, we employ ML to develop a tool to predict electrode voltages for metal-ion batteries using data from the Materials Project database.
One of the challenges associated with a ML approach in materials science is the problem of  finding the proper feature vectors that can accurately represent the
compounds.\cite{Machine1,  bigdata, Burke_ML} In the literature, feature vectors derived from several approaches have been used.\cite{columb, voronoi,  bigdata, Burke_ML}
A simple, yet effective, approach  is   the one derived
from chemical, structural, elemental, and electronic representation of the compounds.\cite{ bigdata}
Here, we  utilize the feature
vectors derived from the chemical properties of compounds and the properties of their elemental constituents in combination with deep neural networks (DNN)\cite{deepLearning, DL1}, support vector machine (SVM)\cite{SVM}, and kernel ridge regression (KRR).\cite{KRR_ref} 
Voltage profile diagrams generated from ML methods are compared to the corresponding DFT based diagrams. 
We also provide a
web-accessible interface that predicts the
voltage of any electrode material (for any metal-ion battery) with minimal basic information and within minutes.
Our work shows that ML models can be employed as an exploratory tool 
to predict  the voltage of electrode materials very efficiently.

\section{Data and Feature Vectors}

Our training data were extracted from the Materials Project Database containing  a total of 4,250 data instances for 3,580  intercalation based electrode materials.\cite{Materials_project,materials_API} We utilized  the Materials Project’s 
application programming interface \textbf{pymatgen} to access the data from the database.\cite{Materials_project,materials_API} Each data instance corresponds to the average voltage calculated for a material in-between two  concentrations of intercalating metal-ion. For some electrodes, average voltages are calculated for multiple concentrations, 
resulting in more data instances in comparison to the number of electrode materials. A total of  3,977 data instances were kept after removing inconsistencies and repetitions.
The dataset that we use contains DFT predicted voltages for several metal ion batteries such as Li, Mg, Ca, Al, Zn, and Y. Nearly 65\% of the
data corresponds to Li-ion battery materials, while data for Ca, Mg, and Zn-ion batteries comprises 10\% each in the database.  Additionally, 4\%
of data instances correspond to  Y and Al-ion batteries. The database provides the average voltage of materials starting from completely
de-intercalated  to fully intercalated (e.g. C$_6$ $\rightarrow{\text{LiC}_6}$, considering C$_6$ as electrode material and Li as the metal) as well as from partially
intercalated to fully intercalated (e.g. Li$_{0.5}$C$_6$ $\rightarrow{\text{LiC}_6}$).

The features used to specify a particular electrode material in our ML models  include
the working ion in the battery (\textit{i.e.} Li or other metal), the concentration of the active metal ion in a given compound, crystal lattice types, and space group numbers.
All other features were obtained from the elemental properties of the atomic constituents involved in a particular electrode. The elemental properties added to the feature vectors are adopted from the work of Ward \textit{et al.}\cite{ward2016general} and
 are listed in the  Supplementary Information (SI). 
This results in 237 features that uniquely represent each compound in the data set.
The components of a feature vector are  diverse in terms of their magnitudes which can range from a few thousands to small fractions. Therefore,
all the input features were normalized for a better and more efficient training of the model that avoids any biased preference of a particular feature
with respect to others based solely on their magnitude.
Normalization was performed by 
effectively scaling all the inputs to be between -1 and 1, and it was 
carried out on-the-fly while training our model and it is not done for target values. 

\begin{figure}[h!]
\begin{center}
\renewcommand{\thefigure}{\arabic{figure}}
\includegraphics[width=0.8\textwidth]{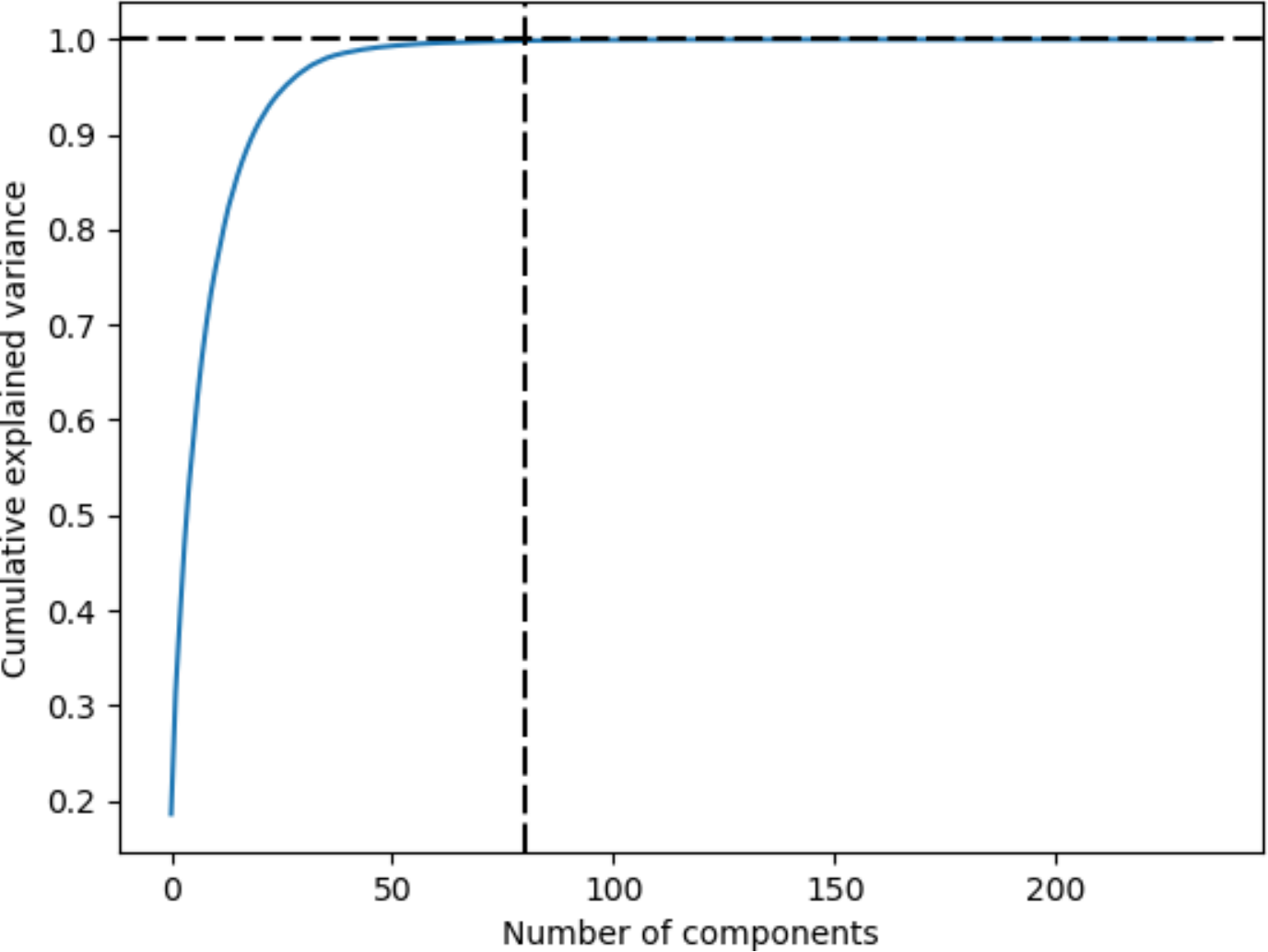}
\caption{Principal component analysis for the feature vectors leads to a reduction of 66\% of the dimensionality.}
\label{PCA}
 \end{center}
\end{figure}

 To systematically remove the redundancy within the features vectors and also to reduce the dimensionality of the features space, we used principal component analysis (PCA)\cite{PCA1, PCA2} algorithms as implemented in the sklearn library\cite{scikit-learn}. 
In Figure~\ref{PCA}, we show the cumulative explained variance as a function of the number of principal components (PC). It shows that only 80 PC are required,
reducing the dimension of feature space significantly. Thus, only these 80 PC were fed into our ML models.

Our data were split into two parts, the training set (T-set, 90\%) and the holdout test set (H-set, 10\%). The T-set is randomly shuffled and used for parameter optimization  with a 10-fold cross validation.  In the 10-fold cross validation, data is randomly divided into 10 equal segments (known as folds), and data in 9 of these folds
are used for training the models and the remaining fold for validating the model. This is repeated 10 times in a way that each of the folds is used as  validation set.
Independent validation of the models was done with the H-set which is not utilized during the training
process. In addition, we also examined the performance of our models on a limited Na-ion batteries data set (Na-set) taken from the literature.\cite{Na-test-set}
To assess the performance of our machine learning models we utilize the mean absolute error (MAE) defined as 
\begin{equation}
    \text{MAE} = \frac{1}{N}\sum_{i}^{N}|Y^{i}_{\text{target}}-Y^{i}_{\text{predicted}}|\,,
    \label{Eq3}
\end{equation}
\noindent
where $Y^{i}_\text{target}$ and $Y^{i}_\text{predicted}$ are the target  predicted values using our models for a given set of parameters for each material $i$, and $N$ is the number of instances in the data set.



\section{Results and Discussion}

\begin{figure}[h!]
\begin{center}
\renewcommand{\thefigure}{\arabic{figure}}
\includegraphics[width=0.8\textwidth]{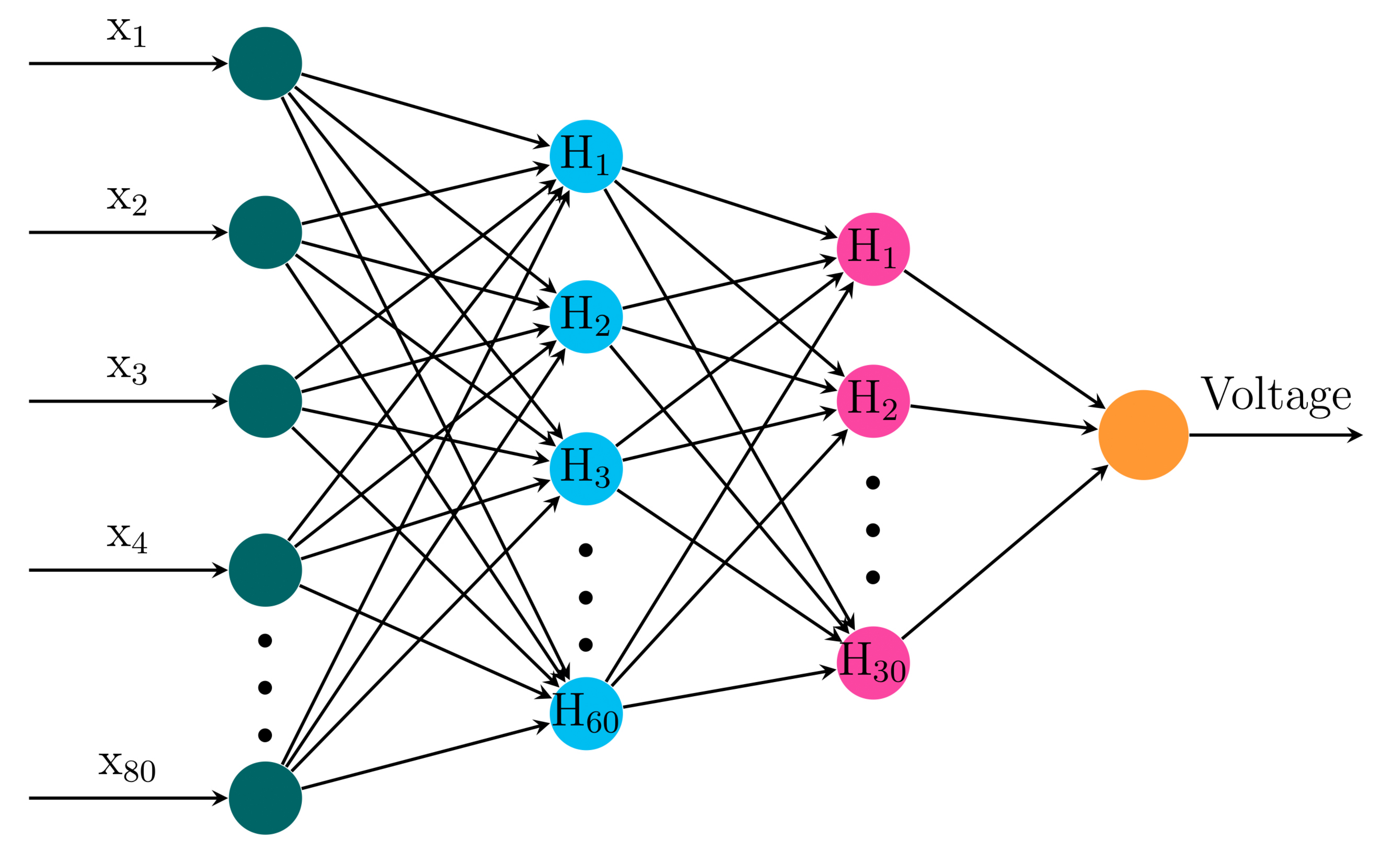}
\label{DNN}
 \end{center}
\caption{ Architecture of the deep neural network used in our work. x$_i$ are the inputs for the input layer and H$_i$ represents nodes in hidden layers. The output from the output layer is the  voltage.}
\label{DNN-MAE}
\end{figure}

With the goal of assessing the performance of different ML algorithms, we have employed three machine learning algorithms: DNN, SVM and KRR. This assessment allows us to select the most robust model for our web-tool.
A 4-layer DNN was used in our work with an architecture shown in Figure~\ref{DNN-MAE}.
To limit overfitting, we added a dropout layer\cite{dropout} after the second and third layer with a rate of 25\% and 10\% respectively. In addition, we used L2 regularization technique in our work.\cite{L2-regularization} 
In the input layer, we have 80 nodes with 60 and 30 nodes in the first and second hidden layer, respectively, and one node in the output layer.
All of these parameters were tuned for optimal performance of the model in the T-set and were chosen optimal when we obtained minimum MAE and when we did not detect signs of overfitting and underfitting.   
We have used mean square error as loss function and MAE as a metric to measure the performance of the model as implemented in the Keras machine learning library\cite{chollet2015keras}. We use ``rmsprop" as an algorithm to optimize the loss function.

\begin{figure}[h!]
  \renewcommand{\thefigure}{\arabic{figure}}
\begin{subfigure}{0.49\textwidth}
\includegraphics[width=1.0\linewidth]{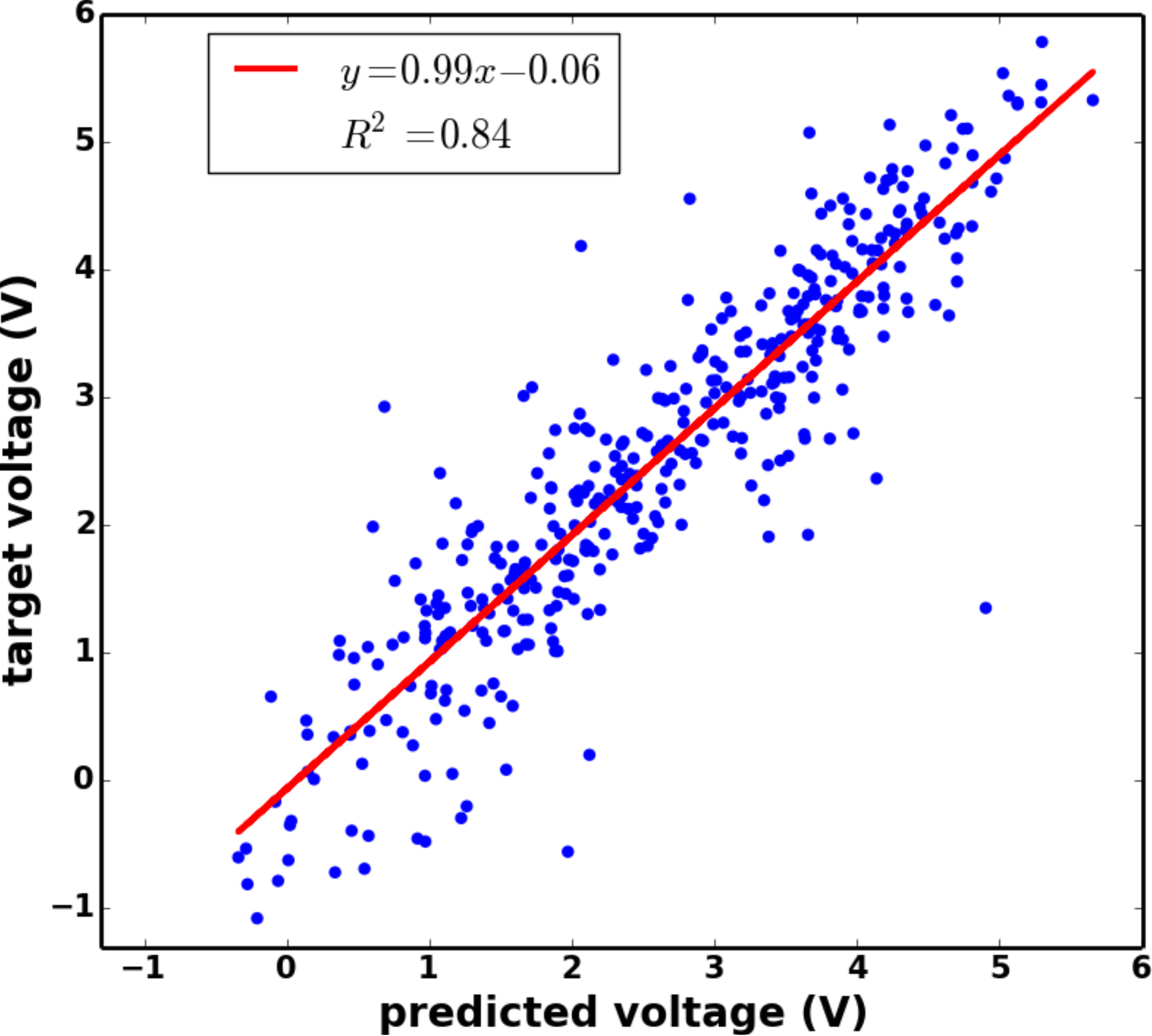}
\caption{\phantom{aaaaaaaaaaaaaaaaaaaaaaaaaaaaaaaaaa}}
\label{SVR}
\end{subfigure}
\begin{subfigure}{0.49\textwidth}
\includegraphics[width=1.0\linewidth]{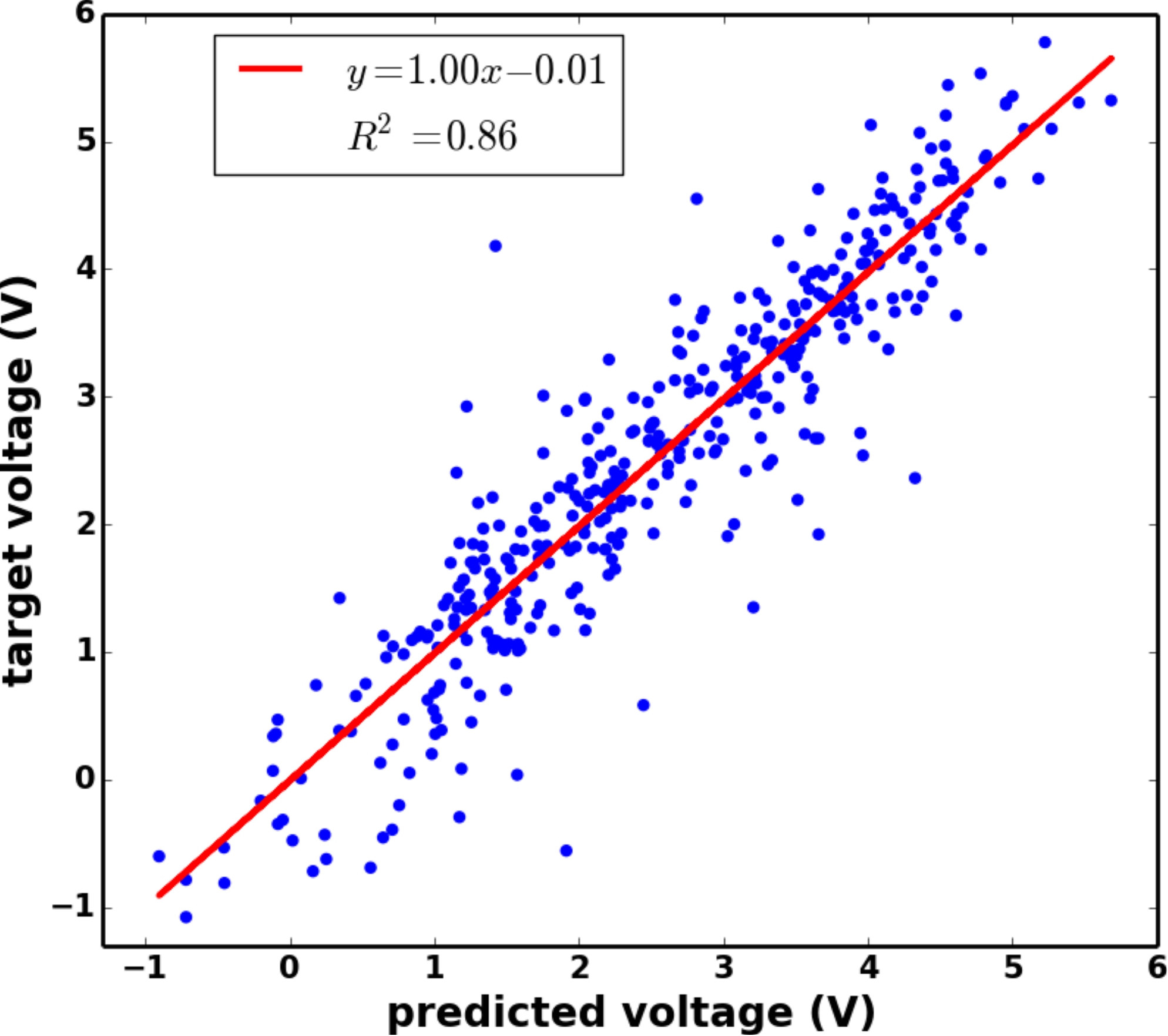}
\caption{\phantom{aaaaaaaaaaaaaaaaaaaaaaaaaaaaaaaaaa}}
\label{KRR}
\end{subfigure}

\begin{subfigure}{0.49\textwidth}
\includegraphics[width=1.0\linewidth]{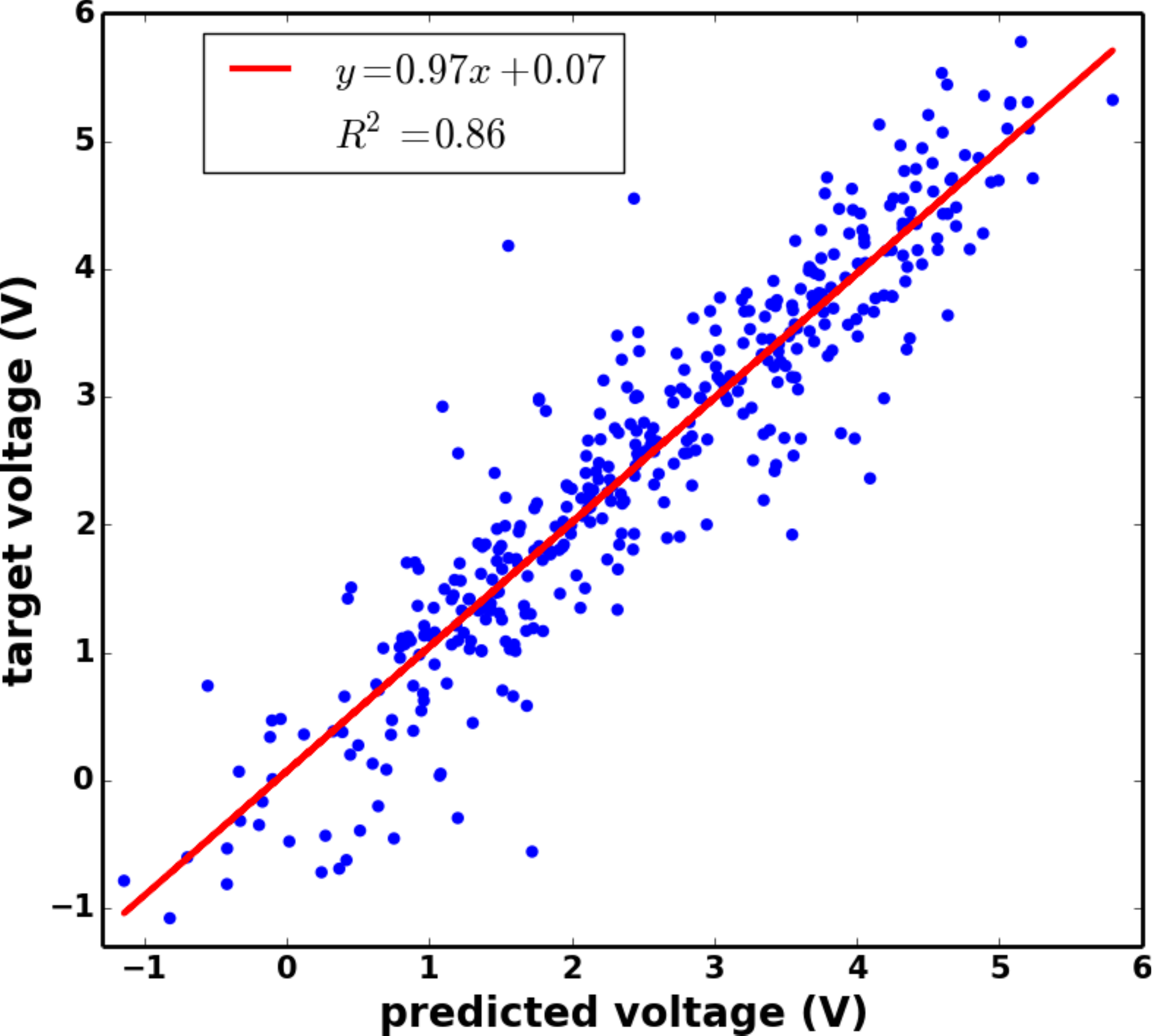}
\captionsetup[subfigure]{oneside,margin={5.0cm,0cm}}
\caption{\phantom{aaaaaaaaaaaaaaaaaaaaaaaaaaaaaaaaaa}}
\end{subfigure}
\begin{subfigure}{0.49\textwidth}
\includegraphics[width=1.0\linewidth]{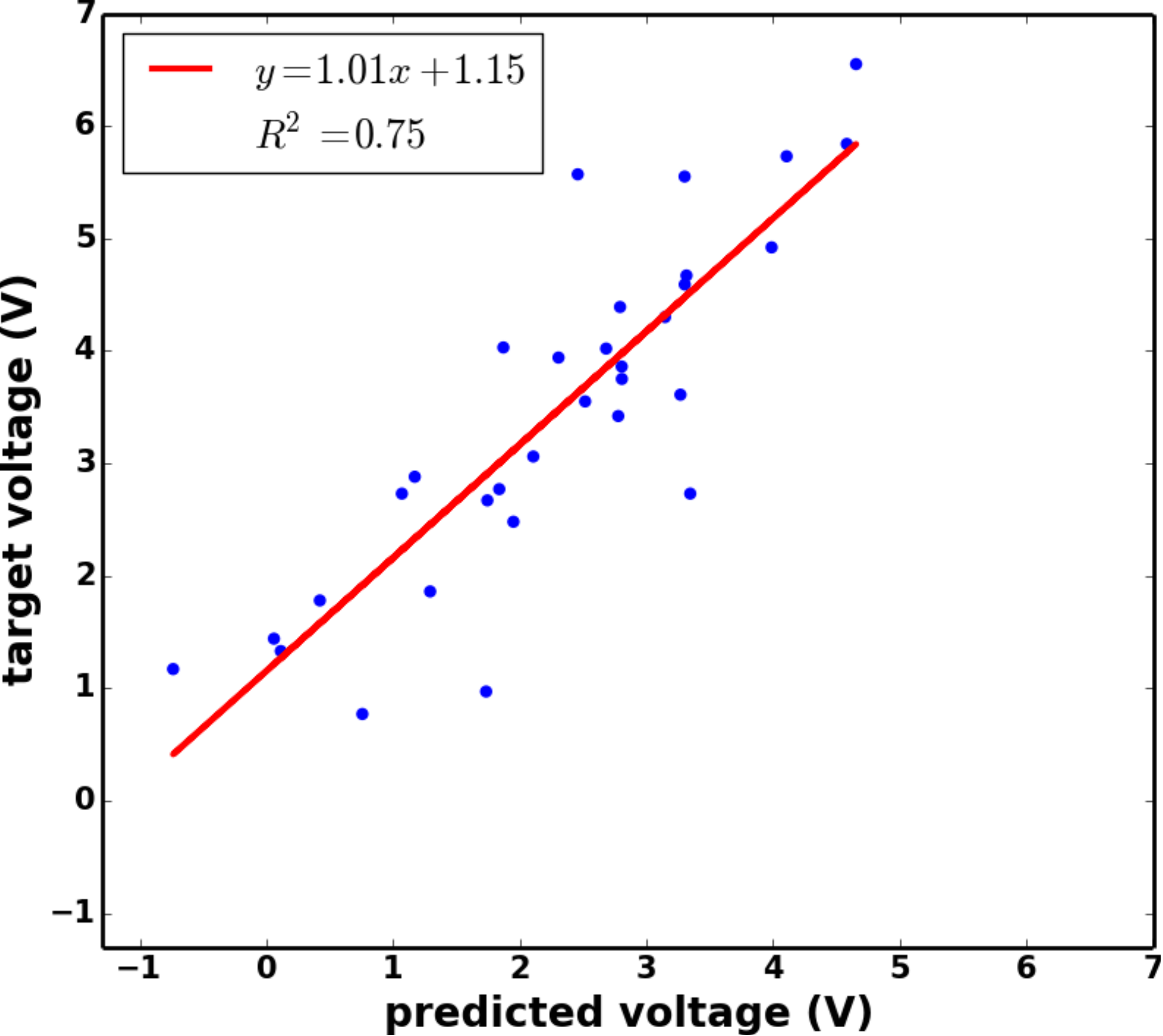}
\captionsetup[subfigure]{oneside,margin={5.0cm,0cm}}
\caption{\phantom{aaaaaaaaaaaaaaaaaaaaaaaaaaaaaaaaaa}}

\end{subfigure}
\label{onNa}
\caption{ Scatter plot showing target vs. predicted voltages  with different ML algorithms used in this work. (a) DNN, (b) SVR, (c) KRR, each on H-set. (d) DNN on Na-set. The best fit equation  ($y=mx + c$) and $R^2$ values for linear fit between  target and ML predicted values are provided as an inset.
}
\label{scatter-plot}
\end{figure}

The MAE and the standard deviation for each fold in the T-set (computed with the optimal parameters) are shown in Table~\ref{elem}.
We obtained similar
MAE in all 10 folds of the DNN model with mean MAE of 0.43 V and standard deviation of $\pm$0.03 V.
We obtained a similar MAE of 0.43 V in the H-set.
The performance of the model in the H-set is shown in Figure~\ref{scatter-plot}(a), where a 
linear relationship between target and DNN predicted values is clearly shown.

\begin{table}[h!]
\centering
\caption{Mean absolute error (V) for each fold of 10-fold cross-validation using deep neural network (DNN), support vector regression (SVR) and kernel ridge regression (KRR). The mean  and the standard deviation of the MAE in these 10 cross-validation sets is given.  In addition,  the MAE for the H-set and Na-set are  provided.
}
\begin{tabular}{l l  l l   }
 \hline \hline
Fold   & DNN       & SVR         & KRR         \\\hline
1      &   0.42    &   0.51      & 0.54        \\
2      &   0.48    &   0.25      & 0.28        \\
3      &   0.42    &   0.26      & 0.27       \\
4      &   0.44    &   0.35      & 0.47       \\
5      &   0.44    &   0.38      & 0.43       \\
6      &   0.42    &   0.62      & 0.71       \\
7      &   0.43    &   0.43      & 0.42       \\
8      &   0.41    &   0.59      & 0.62       \\
9      &   0.45    &   0.53      & 0.57       \\
10     &   0.48    &   0.28      & 0.30       \\\hline
Mean MAE$\pm$standard deviation   & 0.43$\pm$0.03   &      0.42$\pm$0.13   & 0.46$\pm$0.14                                                \\
\hline
 MAE H-set         &  0.43    &        0.40          & 0.39            \\
\hline
 MAE Na-set            &  1.25    &        1.00          & 0.93            \\
 \hline \hline
\end{tabular}
\label{elem}
\end{table}



\begin{figure}[h!]
\begin{subfigure}{0.49\textwidth}
\includegraphics[width=1.0\linewidth]{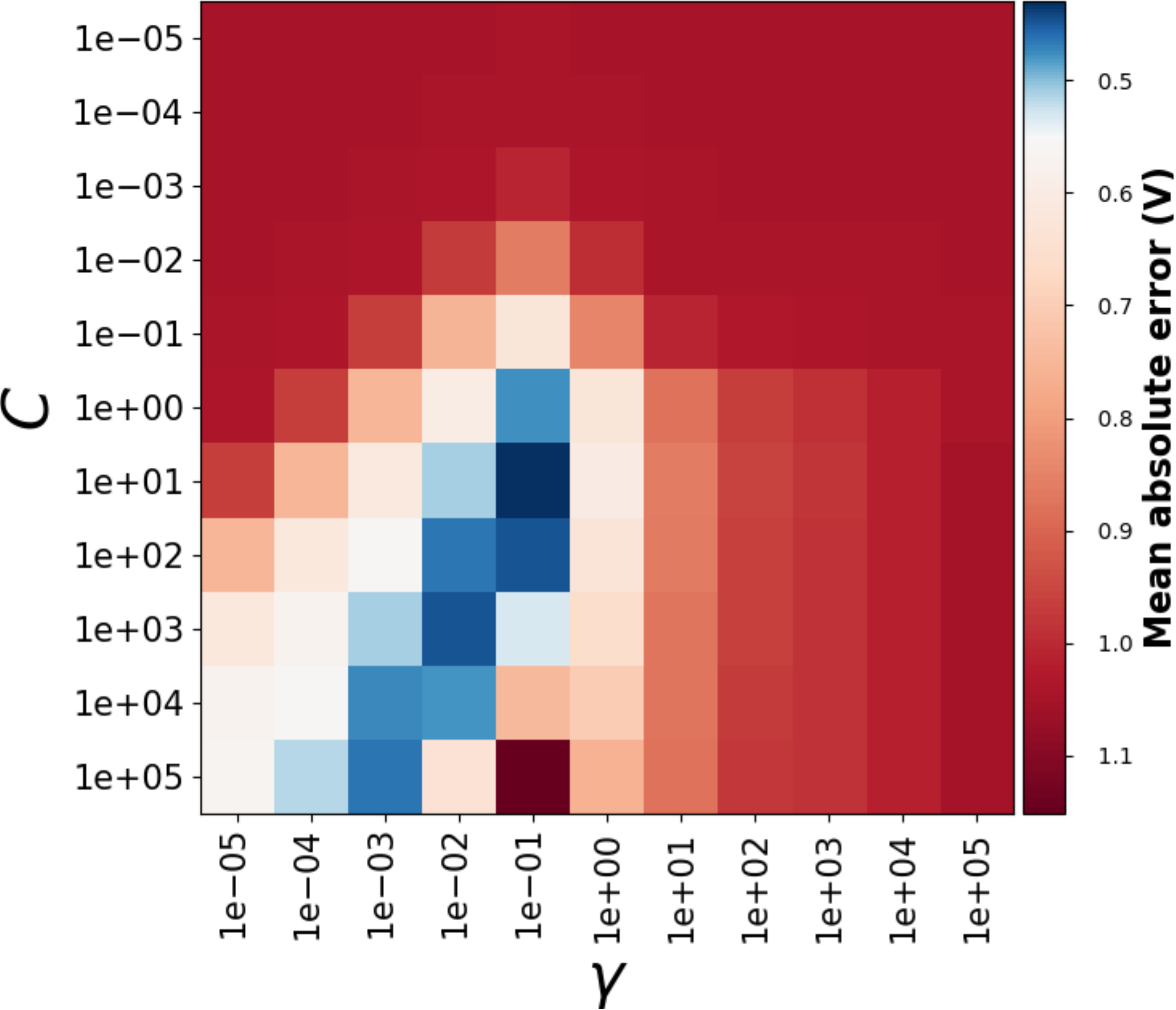}
\captionsetup[subfigure]{oneside,margin={5.0cm,0cm}}
\caption{\phantom{aaaaaaaaaaaaaaaaaaaaaaaaaaaaaaaaaa}}
\end{subfigure}
\begin{subfigure}{0.49\textwidth}
\includegraphics[width=1.0\linewidth]{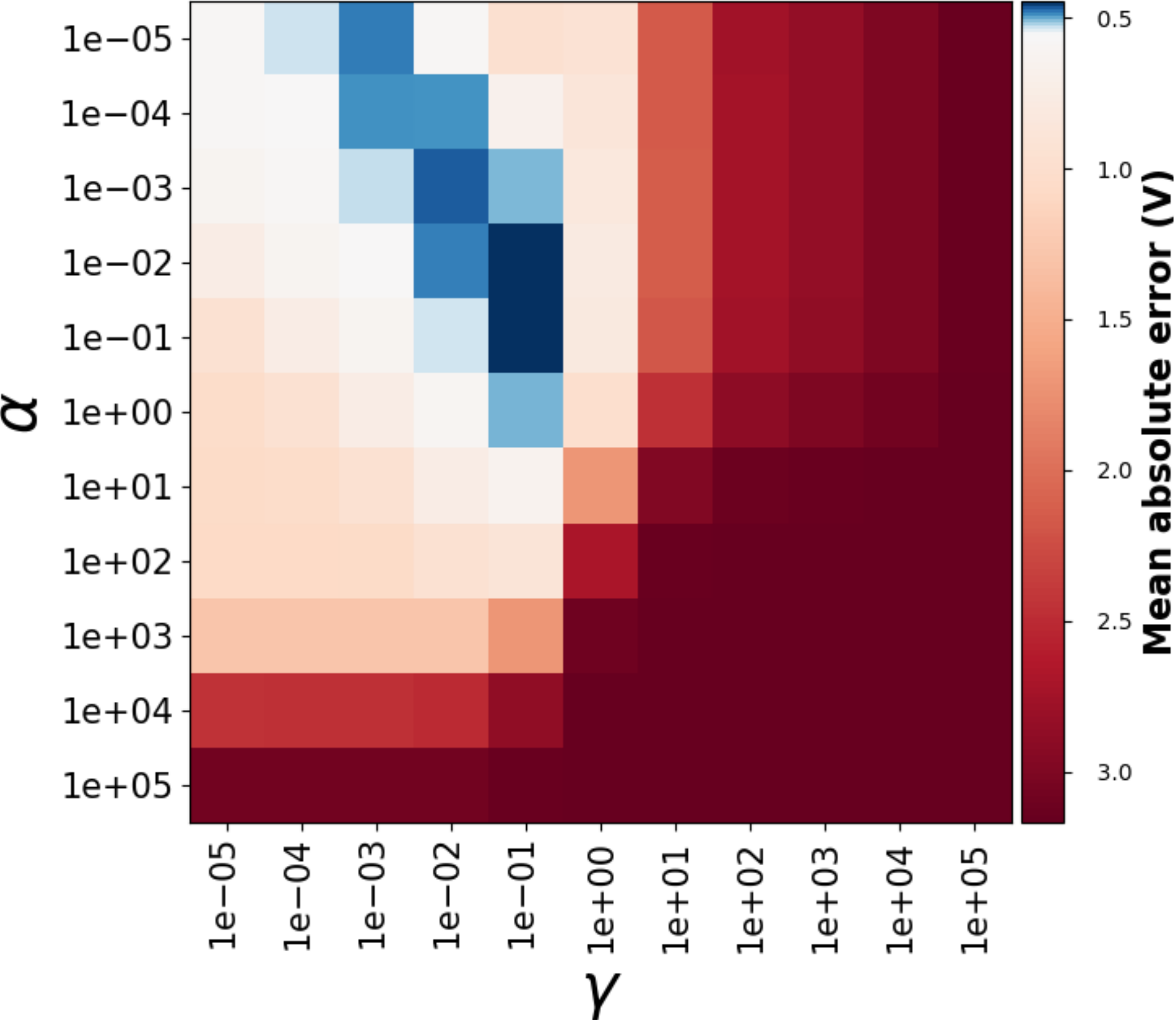}
\caption{\phantom{aaaaaaaaaaaaaaaaaaaaaaaaaaaaaaaaaa}}
\end{subfigure}
\caption{ Color maps showing the tuning of  (a) $C$ and $\gamma$ parameters for support vector regression and (b)  $\alpha$ and $\gamma$ parameters for  kernel ridge regression. The color bars display the  mean absolute error.}
\label{grid_Search}
\end{figure}

We compared the performance of DNN with another machine learning model, SVM.
When used for a regression problem, SVM  is known as a support vector regression (SVR). 
SVR is a kernel-based
regression technique known for its robust performance in complex data representations. 
It works by mapping
 non-linearly separable data in real space to higher dimensional space $via$ a kernel function.
We have used the radial basis function (RBF) kernel for this work.
In addition, SVR  depends on two  important parameters ($C$- and $\gamma$) that control the quality
of the result. These parameters were tuned by using the grid search algorithm of sklearn. We varied $C$  and
$\gamma$ logarithmically in between 10$^{-5}$ and 10$^5$ each. For each of the possible combinations of $C$ and  $\gamma$, SVR computations
were performed using a 10-fold cross-validation to calculate the mean of the MAE.  The parameter space and  MAE encoded as color, obtained from a grid search is shown in Figure~\ref{grid_Search}(a).
We determined optimal values of $C$ and $\gamma$ such that they yield the minimum MAE from the grid search, which corresponds to an optimal combination of  $C=10.0$ and $\gamma=0.1$ for which we obtain a mean MAE of 0.42$\pm0.13$~V in the T-set. 
Further refining the grid in the small range around the optimal values of $C$  and $\gamma$  does not qualitatively influence our results.
The MAE for each fold of 10-fold cross-validated T-set obtained with these tuned parameters and the RBF kernel are given in  Table~\ref{elem}.
The performance of SVR for the H-set is shown in Figure~\ref{scatter-plot}(b).

In addition to SVR, the performance of DNN was compared with another kernel based ML algorithm known as kernel ridge regression (KRR). Similarly to SVR, we used a grid search technique to find the optimal parameters $\alpha$ and $\gamma$ by varying each between 10$^{-5}$ and 10$^5$, which is shown in Figure \ref{grid_Search}(b). In this case we obtained an optimal value of $\alpha$ and $\gamma$ as 0.01 and 0.1, respectively, for which KRR yields a mean MAE of 0.46$\pm$0.14~V in T-set. 
The MAEs in the 10 folds of T-set for the KRR model are given in Table~\ref{elem} 
alongside the values obtained from DNN and SVR.  
Finally, we checked the performance on the H-set, which  is shown in Figure~\ref{scatter-plot}(c). We obtained a MAE of 0.39~V on the H-set which is in agreement to what we observed using SVR (0.40~V). 
These MAEs for the H-set depend only marginally on the ML algorithms,
although, with SVR and KRR, there are minor  deviations between the folds related to the kernel based nature of both SVR and KRR: as we move between the folds, the data  might not be that  separable in  folds when mapped  using the kernel functions.


The performance of our ML models on the H-set is shown in Figure~\ref{scatter-plot}(a-c).
We show in these scatter plots that for lower voltage ranges the models perform slightly worse than for the higher voltage range. We attribute this behavior to the limited amount of data available for training the model in this range. Such a limited amount of data might not be sufficient to capture the complexity of the patterns needed to predict voltages properly. We believe that the performance of the model in the low voltage region could be improved by adding more data points  within this range to the T-set.

For each model, we also show in Figure~\ref{scatter-plot}(a-c) the best-fit linear equation and the $R^2$ value  as a measure of goodness of fit between the ML predicted and the target values in the hold out test set. 
In the ideal case, the equation of best fit should be $y=x$ whereas $R^2$ should be one.
 With our models, we observed 
this  relationship between the target and the ML predicted values. In addition, the good performance of our models  is also reflected in  the  $R^2$ values. We obtained reasonable $R^2$ values of 0.84, 0.86, and 0.86 with DNN, SVR, and KRR respectively. 


We note that although our ML-models perform well and consistently, a more challenging test for the models would be achieved by gauging the performance of our models on a completely new data set. 
To this end, we checked the performance of the models on the Na-set. This data set is taken from the work of Zhang \textit{et al}.\cite{Na-test-set} and contains voltage information of 32 Na-battery electrode materials with voltages in the range of 0.7 - 6.5 V. 
This set of materials was obtained
by screening the entire Materials Project database for Na-based layered materials and then performing DFT calculations on the selected set of materials suitable as battery electrodes.  
The Na-set differs from the  original  T- and H-set as it contains the voltage for Na-based electrode materials (for which our model is not trained). 
Thus, this is an excellent test to examine the robustness and the transferability of our ML models.
In Table~\ref{elem}, we list the MAE obtained for the Na-set with each of the ML models. 
In addition, 
  in Figure~\ref{scatter-plot}(d)
we show the scatter plots obtained with DNN. We observed that the model does not perform as well for Na-set as it does during cross validation for the T- and  
H-sets. 
We obtained a MAE about 0.8~V larger than in the case of the T- and H-set using DNN. Despite  the large MAE, the linear relationship between the predicted and targets values is still preserved with our ML models. However, the poorer performance of the models for this test set is  reflected in the relatively small $R^2$ value as well as the best fit line with large intercept. 
 We identified two main reasons that could explain the comparatively poorer performance of our models in the Na-set. First, our model is not trained for Na-ion electrodes due to the lack of data in the databases for Na. 
 Secondly, the Na-test set contains only 32 materials which can affect the statistical performance of the ML models.


\subsection{Li-only Data}

Our data set is diverse as it contains 6 different metal-ions (Li, Ca, Mg, Zn, Al, and Y) battery electrodes in a relatively small amount of data. It is interesting to see how the ML models perform on a specific metal-ion battery data. As a representative case, and due to the availability of a relatively large proportion of data for Li-ion electrodes, we trained our ML models only on Li-only battery data (Li-set). The model parameters are tuned again with Li-only data using the same protocol as we used for the entire data (parameters are provided in the SI). 
The MAE obtained from all three ML models in
 10-fold cross validated T-set, H-set, and Na-set
are given in Table \ref{Li-only}. We note that T-set and H-set used in this case are 90\% and 10\% of the Li-only data. 
The corresponding scatter plots in the H-set are provided in the SI.


\begin{table}[h!]
\centering
\caption{Mean absolute error (V) for each fold of 10-fold cross-validated T-set, H-set, and Na-set with DNN, SVR and KRR  along with  mean  and standard deviation in MAE for the models trained on Li-set. Note that the T-set and H-set here are taken from the Li-only data set.
}

\begin{tabular}{l l  l l   }
 \hline \hline
Fold   &   DNN     &   SVR       & KRR        \\\hline
1      &   0.44    &   0.65      & 0.64       \\
2      &   0.45    &   0.37      & 0.45       \\
3      &   0.53    &   0.30      & 0.34       \\
4      &   0.43    &   0.36      & 0.37       \\
5      &   0.48    &   0.43      & 0.42       \\
6      &   0.50    &   0.79      & 0.78       \\
7      &   0.48    &   0.47      & 0.46       \\
8      &   0.43    &   0.39      & 0.40       \\
9      &   0.48    &   0.68      & 0.73       \\
10     &   0.46    &   0.57      & 0.53       \\\hline
Mean MAE$\pm$standard deviation   & 0.47$\pm$0.03   &      0.50$\pm$0.15   & 0.51$\pm$0.15                                                \\
\hline
 MAE H-set      &  0.42    &  0.44   &    0.44         \\
\hline
 MAE Na-set         &  0.70    &        0.62          & 0.70            \\

 \hline \hline
\end{tabular}
\label{Li-only}
\end{table}

 \begin{figure}[h!]
 \begin{center}
 \renewcommand{\thefigure}{\arabic{figure}}
\includegraphics[width=0.65\textwidth]{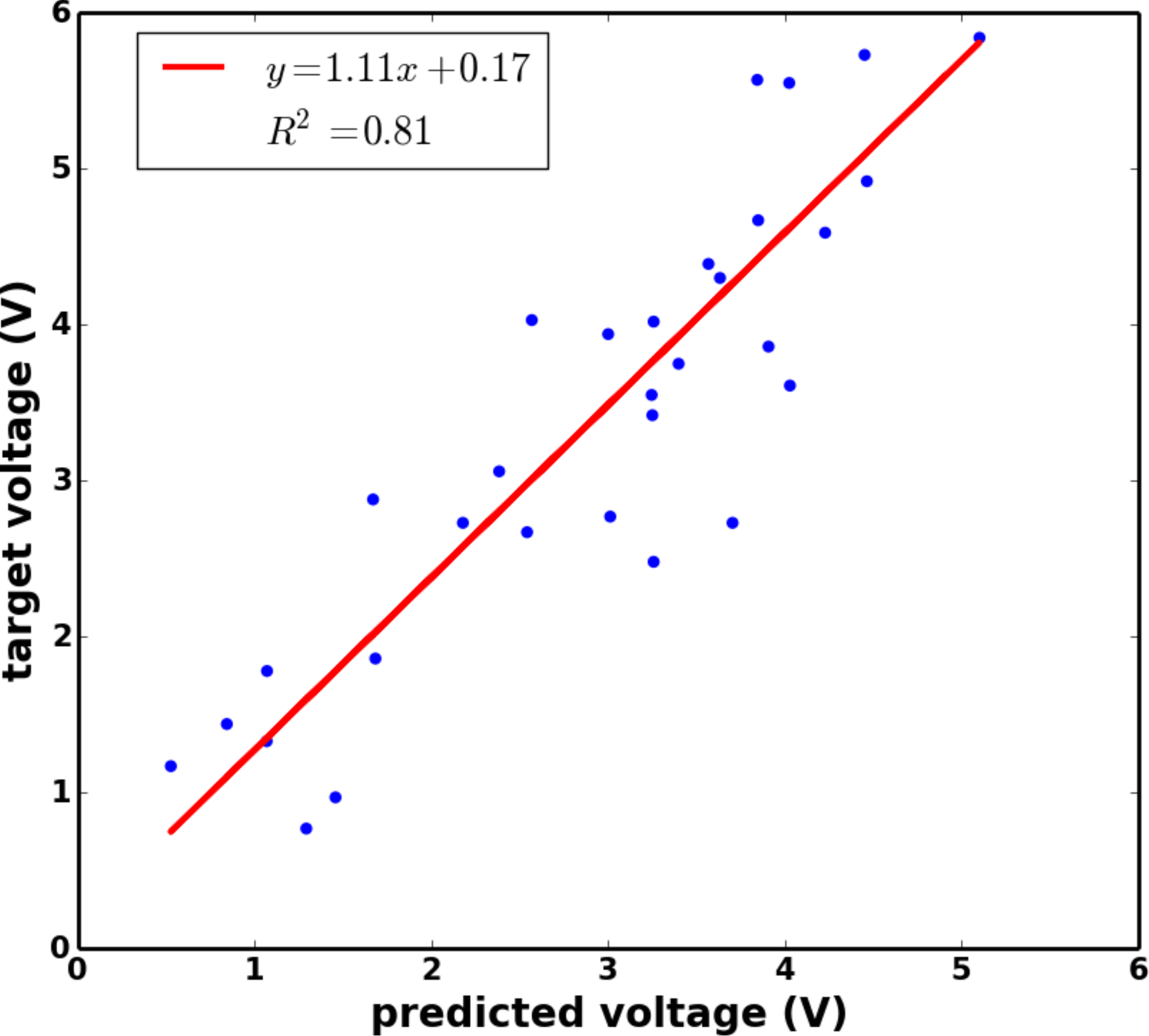}
\caption{ Scatter plot showing the target $vs.$ the predicted values (V) with DNN trained on Li-set and applied to Na-set which is taken from the work of Zhang \textit{et al.}\cite{Na-test-set} (other scatter plots are available in the SI).}
\label{na_NNN}
 \end{center}
\end{figure}

Although the performance of the models in the T-set and H-set, as quantified by the MAE, is similar to the one obtained considering the entire data set,  
the performance on the Na-set is improved in comparison to the models trained on the entire data set (see Table \ref{elem}). 
Such an improvement might be due to the more coherent data for training the models as well as the  similarity between Na-ion and Li-ion electrochemistry.
In  the case of models trained only on Li-data, a MAE of 0.62-0.70~V is obtained for the Na-set with all three models,  which is slightly larger than what we obtained for the other test sets. This larger error is somewhat expected, and still promising, considering that the model is not trained for Na battery electrode materials and  the small number statistics effect in the Na-set. Nonetheless, these results show that our ML models are transferable among  different metal-ion batteries.

In Figure~\ref{na_NNN}, we show the scatter plot obtained by using DNN on the Na-set. The performance of DNN on this new test set is similar to what we observed for the original Li-only training and test sets. The ML predicted voltage values follow a linear trend with respect to the target voltage. The improved performance on Na-set can also be seen from the best fit equation ($y = 1.11~x+0.17$, compared to the entire data best fit equation $y = 1.01~x+1.15$ on Na-set).
Additionally, we observe an  improvement in the $R^2$ value with the model trained only on Li data batteries for the Na-set.
Similar results were obtained with SVR and KRR (the corresponding scatter plots are provided in the SI).

\subsection{Voltage Profiles}

\begin{figure}[h!]
\includegraphics[width=0.8\linewidth]{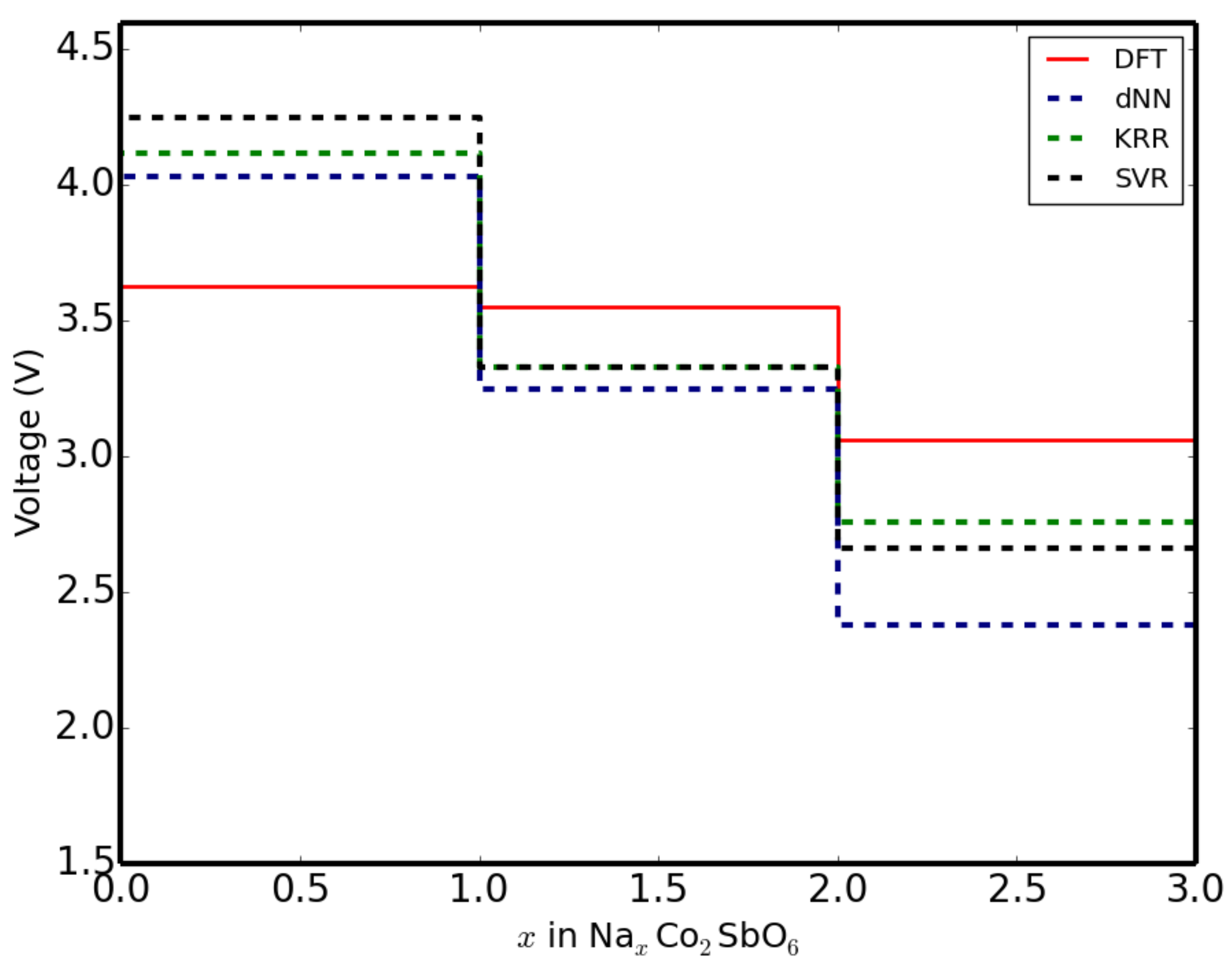}

\caption{ Voltage profile diagram obtained from different ML models and DFT for  Na$_x$Co$_2$SbO$_6$. The DFT values are taken from the work of Zhang \textit{et al.}\cite{Na-test-set}}

\label{vprofile}
\end{figure}

The ML models we developed in this work can  be used to generate the voltage profile diagram for electrode materials.  These diagrams provide an estimate of the voltage range that an electrode can supply when used in a battery. 
To obtain these diagrams one needs to evaluate voltages at different intercalation concentrations. Our ML method offers a fast alternative to DFT methods estimate this voltages.  As an example, in Figure~\ref{vprofile} we show the voltage profile diagram of the Na-based electrode material 
Na$_x$Co$_2$SbO$_6$, obtained from ML models and compare it with the one obtained using DFT, as  reported by Zhang \textit{et al.}\cite{Na-test-set}
In the voltage profile diagrams, we observe that as the concentration of Na-ion increases, the DFT voltage decreases gradually. The same pattern is observed with the DNN predicted voltage in the electrode material considered. This trend is also reproduced by the other two machine learning algorithms (SVR and KRR) as shown in Figure~\ref{vprofile}. 

\subsection{Application: Prediction of New Electrodes}


The potential of our ML models to propose new electrode materials for Na- and K-ion batteries is assessed by studying  candidate materials  that are  known for Li-ion but not necessarily explored for Na- and K-ion batteries. In the literature, some of the materials used as electrode for Li-ion have also been studied for their use as electrodes in Na-ion batteries.
For instance, Ceder \textit{et al}.\cite{Exp-ML1} presented a comparative study for a set of selected cathode materials in terms of the voltage they offer when used as Li- and Na-ion battery electrodes. They found that a material provides on average 0.18-0.57~V lower voltage when used as electrode for Na-ion battery than the corresponding Li based electrodes, in agreement with experimental observations in some of these materials. 

Although the generalization of their conclusion for all possible Li and Na based electrode materials is not documented, the trend observed from their study 
is encouraging as it implies that electrode materials that have been known over the last two decades for their use in Li-ion batteries can potentially work as electrodes in Na- and K-ion batteries operating at lower voltages. The majority of Li-based electrode materials existing in the MP-database are not explored for Na- and K-ion battery electrodes. A quantitative estimate of their performance for Na- and K-ion batteries is necessary to build a robust and cheaper alternative to Li-ion batteries. 
Here, we use our ML tool to examine the performance of Li-based electrode materials found in the MP-database for Na and K by replacing the intercalating Li-ion with Na- and K-ions. 
We utilize the same stoichiometry and crystal structure for Na and K intercalated material as in the parent Li-based material. The goal of this exercise is to propose new cathodes for Na- and K-ion batteries, predict their voltage and study the correlation of predicted voltage with respect to that of Li-ion. 

We first verify the conclusion of Ceder \textit{et al.}\cite{Exp-ML1} by  taking a few materials from their work and calculating the voltage of these materials for Li-, Na-, and K-ions using our DNN ML model. 
These results are shown in Table~\ref{Li-Na-K}, where we include DFT and experimental values when available.
Our ML model reproduced the trend observed from DFT calculations. ML predicted voltages for Na-ion materials are on average 0.30 V smaller than the corresponding Li values. This is comparable to the 0.40 V shift observed from DFT calculations for the same materials. The voltages predicted for K-ion materials are even smaller than the corresponding voltages in Na-ion materials. Overall, average voltages follow the trend Li$>$Na$>$K. 

\begin{table}[h!]
\centering
\caption{ML predicted voltage (V) for Li, Na and K-based electrodes. Corresponding DFT and experimental values (when available) are provided in parenthesis. DFT voltages, experimental references, and materials were taken from the work of Ceder \textit{et al.}\cite{Exp-ML1}
}

\begin{tabular}{l l  l l   }
 \hline \hline
Material                    &   A=Li            &   A=Na             & A=K        \\\hline
ACoO$_2$                    &   3.56 (3.99, 4.10)     &   3.40 (3.48, 2.80)      &  3.26      \\
ANiO$_2$                    &   3.81 (3.82, 3.85)     &   3.61 (3.31, 3.00)      &  3.12      \\
ATiO$_2$                    &   1.96 (1.94)           &   1.79 (1.37, $>1.50$)   &  1.51      \\
ATiS$_2$                    &   1.71 (1.82)           &   1.66 (1.64)            &  1.53      \\
AFePO$_4$                   &   3.36 (3.45, 3.50)     &   2.95 (3.08, 3.00)      &  2.39      \\
AMnPO$_4$                   &   3.66 (3.89, 4.10)     &   3.30 (3.59)            &  2.69      \\
ACoPO$_4$                   &   4.03 (4.64, 4.8)      &   3.53 (4.19)            &  2.85      \\
ANiPO$_4$                   &   4.50 (5.06, 5.3)      &   4.12 (4.58)            &  3.28      \\
A$_3$V$_2$(PO$_4$)$_3$      &   3.54 (3.13, 3.8)      &   3.11  (2.94)           &  2.59     \\

 \hline 
\end{tabular}
\label{Li-Na-K}
\end{table}

The voltage of new electrode materials based on Na- and K-ions is shown in Figure~\ref{K,Na-Li}(a), where we show the scatter plot of the ML predicted voltages of Na-based electrodes against the DFT predicted voltage of Li-based electrodes for the same materials. For the majority of the materials studied, the conclusion of Ceder \textit{et al.} remains valid: 
ML predicted voltages for Na-based electrodes are, in general, smaller than the corresponding Li ones.  We obtained a mean difference and mean absolute difference between DFT Li-voltages and ML-predicted Na-voltages of 0.36~V and 0.56~V, respectively. 

\begin{figure}[h!]
\begin{subfigure}{0.49\textwidth}
\includegraphics[width=1.0\linewidth]{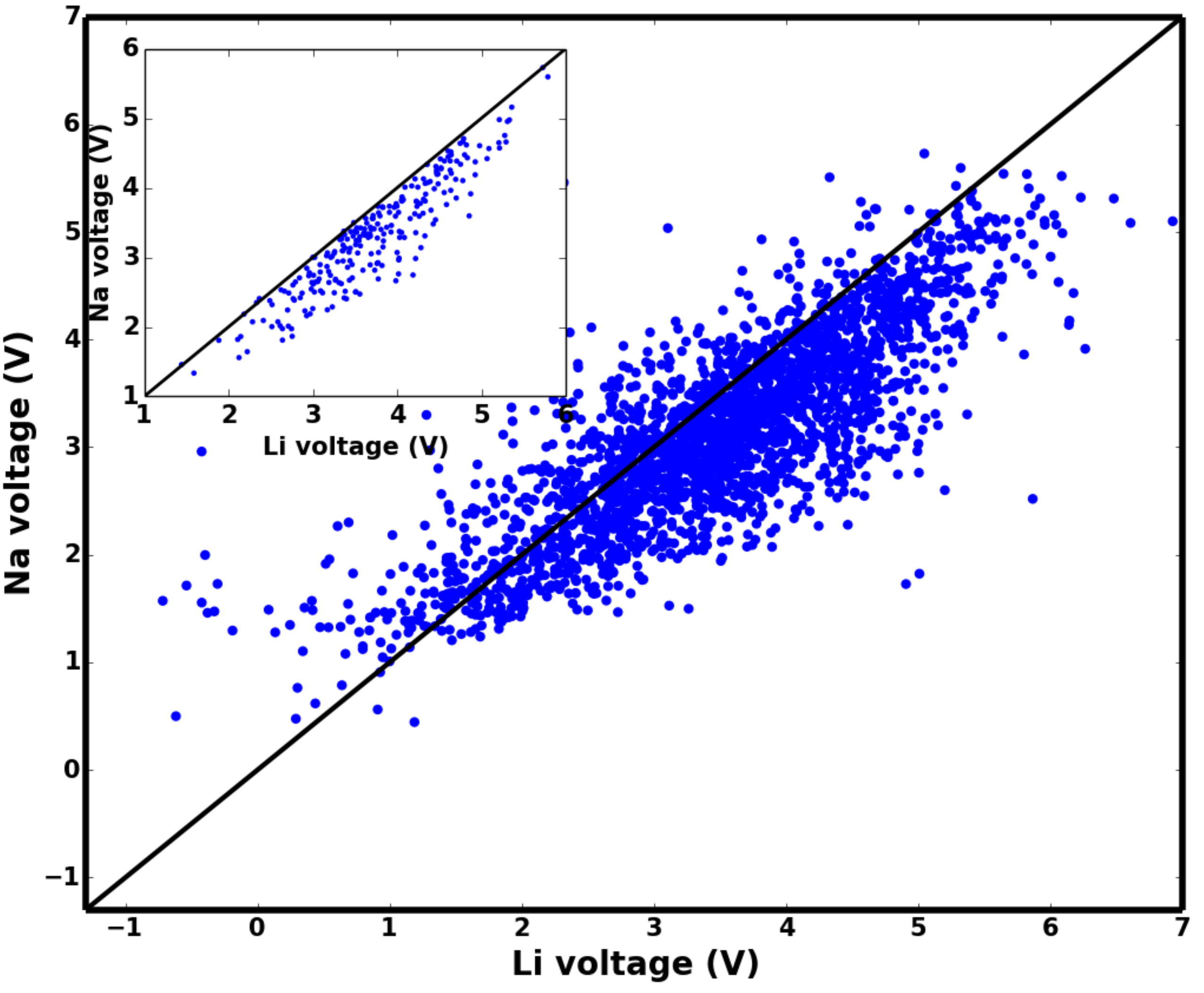}
\captionsetup[subfigure]{oneside,margin={5.0cm,0cm}}
\caption{\phantom{aaaaaaaaaaaaaaaaaaaaaaaaaaaaaaaaaa}}
\end{subfigure}
\begin{subfigure}{0.49\textwidth}
\includegraphics[width=1.0\linewidth]{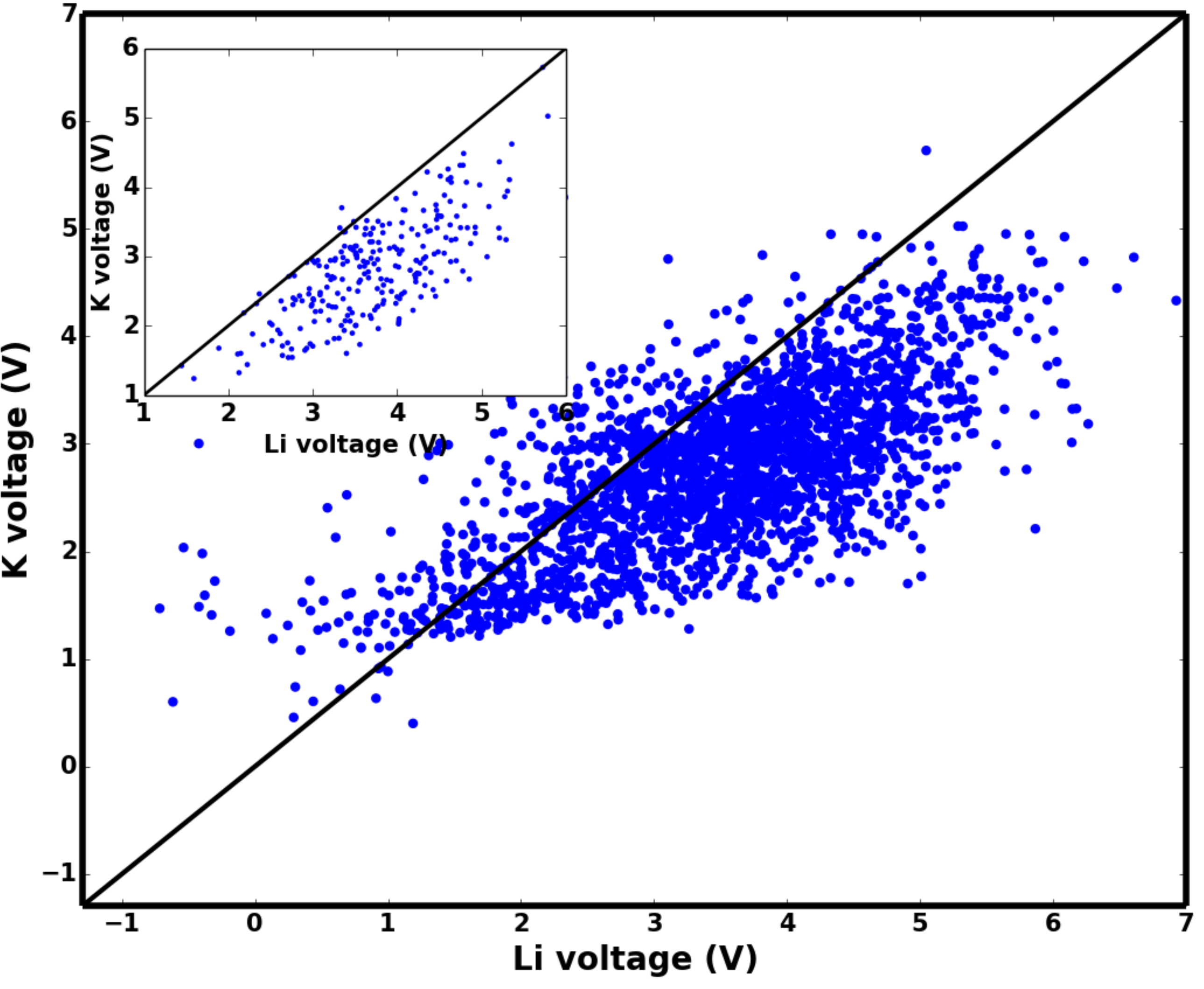}
\caption{\phantom{aaaaaaaaaaaaaaaaaaaaaaaaaaaaaaaaaa}}
\end{subfigure}
\caption{ Scatter plots showing ML predicted voltages for electrode materials obtained by replacing Li ions with (a) Na ions and (b) K ions with respect to the corresponding DFT Li-voltages. The inset shows the ML predicted voltages for Na and K-ion in the H-set of Li-only materials with the corresponding ML-predicted voltages for Li-ion. The $y=x$ line is drawn as a reference for the eye. 
}
\label{K,Na-Li}
\end{figure}


For several electrodes, however, a larger voltage is predicted by our model for Na-based electrodes compared to the corresponding Li-based DFT values. This overestimation might arise because we are comparing ML-predicted voltages of Na-ion against DFT voltages of Li-ion. A better comparison for such correlation, perhaps, should consider ML-predicted voltages for both ions. To this end, we compared the ML-predicted voltage for the H-set used in the Li-only case for both Li- and Na-ion. The resulting scatter plot is shown in the inset of  Figure~\ref{K,Na-Li}(a). In this case, all the predicted voltages for Na-electrodes are smaller than the corresponding ones for Li-electrodes. The corresponding scatter plots for DFT Li-voltage and ML-predicted voltage for Na and K-based electrode materials in the H-set is provided in the SI (Figure~S4). We obtained a mean deviation 
of 0.42~V in the H-set, which is consistent with the trend observed by Ceder \textit{et al.}\cite{Exp-ML1}

A similar analysis was performed for K-based electrodes. We observed that voltages calculated for K-ion are even smaller than in Na- and Li-ions. The corresponding scatter plot is given in Figure~\ref{K,Na-Li}(b). 
ML predicted K-voltages are in general smaller by 0.73~V compared to DFT Li-voltages.  With respect to Li-electrodes in the H-set, ML predicted K-voltages are smaller on average by 0.91~V. The ML-predicted voltages for Na- and K-based electrodes are provided in the SI along with the DFT voltages for the corresponding Li-electrodes.

With this analysis, we have identified a vast number of materials as well as their corresponding voltages that could work as suitable electrodes for Na- and K-ion batteries. From this first screening, researchers could further refine a set of materials in terms of structures, maximum metal stoichiometries, and voltages using DFT methods and, ultimately, experiments.

\subsection{Web Tool for Voltage Prediction}

We  implemented a publicly available  web-accessible tool for the battery community that can be used to predict the voltage of any novel  electrode with minimal basic information typically within a minute. 
The only information required to predict the voltage are the stoichiometry of the intercalated material at high ion concentration, the stoichiometry of the material at low ion concentration, the type of metal-ion battery, the fraction of metal ions corresponding to the stoichiometry of the material at high ion concentration, the crystal lattice type, and the space group.  
This information is readily available for any material with a crystalline structure,  making our tool efficient as well as remarkably easy to use. We used our most robust model, DNN, as a back-end ML model in our online tool for voltage prediction. Our tool can be used to predict the average voltage of Li-, Na-, K-, Mg-, Ca-, Zn-, Al- and Y-ion battery electrodes. 
Li-only trained model is used for prediction of voltage for battery electrodes based on Li-, Na-, and K-ion whereas the entire data set trained model is used for rest of ions. This tool can be freely accessed at  \url{http://se.cmich.edu/batteries}.

\begin{table}[!ht]
\centering
\caption{Table showing the DNN predicted vs experimental average voltage (V) for different electrodes.}

\begin{tabular}{l l  l   lll } 
 \hline \hline
Electrode             &   ML      &   Experiment \vline & Electrode                   &   ML      &   Experiment         \\\hline
NaMnO$_2$             &   2.98    &   2.75$^a$          & LiFePO$_4$                  &   3.36    &   3.50$^b$         \\
NaCoO$_2$             &   3.40    &   2.80$^b$          & LiNiO$_2$                   &   3.81    &   3.85$^b$         \\
NaTiO$_2$             &   1.79    &$>$1.50$^b$          & NaFe$_{0.5}$Co$_{0.5}$O$_2$ &   3.58    &   3.14$^d$                \\
NaNiO$_2$             &   3.61    &   3.00$^b$          & K$_{1.6}$Na$_2$Mn$_3$O$_7$  &   2.74    &   2.20$^e$         \\
LiCoO$_2$             &   3.60    &   4.10$^b$          & Mg$_2$Mo$_6$S$_8$           &   1.09    &   1.30$^f$      \\
NaFePO$_4$            &   2.95    &   3.00$^b$          & Mg$_{0.55}$TiSe$_2$         &   1.63    &   1.45$^g$               \\
Na$_4$MnV(PO$4$)$_3$  &   3.42    &   3.00$^c$          & MgMoO$_3$                   &   2.20    &   2.25$^h$       \\

 \hline \hline     
\multicolumn{6}{l}{ 
$^a$ taken from Ref.~\citenum{NaMnO2},} \\
$^b$ taken from Ref.~\citenum{Exp-ML1}, \\
$^c$ taken from Ref.~\citenum{Na4MnV(PO4)3}, \\ 
$^d$ taken from Ref.~\citenum{Na1}, \\
$^e$ taken from Ref.~\citenum{Na2Mn3O7},   \\
$^f$ taken from Ref.~\citenum{Mg2Mo6S8}, \\
$^g$ taken from Ref.~\citenum{Mg0.55TiSe2}, \\
$^h$ taken from Ref.~\citenum{MgMoO3}.   \\
\end{tabular}
\label{ML-Exp}    
\end{table}

Knowing the performance of the web-tool in comparison to experimental reference might be of interest to the  community exploring the electrodes material experimentally.  As a proof of concept,  we  compare  predicted average voltages with  experimental average voltages. We selected few electrode material from the literature\cite{NaMnO2, Exp-ML1, Na4MnV(PO4)3, KFeSO4F, Na2Mn3O7, Mg2Mo6S8, Mg0.55TiSe2, MgMoO3, KTiS2} and calculated their average voltage using the web-tool. These results are summarized in Table~\ref{ML-Exp} where we observe a good agreement between the predicted voltages and the corresponding experimental values.

\section{Conclusions}

In conclusion, we used deep neural network, support vector regression, and kernel ridge regression to machine learn the voltage of electrode materials in a given metal-ion battery utilizing DFT data extracted from the Materials Project Database. The performance of the models is gauged by comparing the mean absolute error between the training set, the holdout test set, and data from the literature. 
Our results indicate that  ML models reproduce DFT trends and thus can be used to   explore electrode materials in terms of voltages.  
This methodology is fast compared to DFT calculations and can be used to guide experiments seeking to develop novel materials for battery applications or to perform a quick screening before starting synthesis procedures.  
Using our models, we propose nearly 5,000 electrode materials for Na- and K-ion batteries. Further improvement in the performance of the model might be necessary for routine application of ML algorithms for the prediction of voltage of electrode materials. Such improvements might include, but are not limited to, using other  machine learning algorithms, using more data, or exploring other ways of feeding intercalation reactions to the ML models. 
In addition, for the robust training of ML models in the future,  DFT calculations records  of electrode materials in all voltage ranges should be kept  in the databases (even if a material is not suitable as a battery electrode)  since negative data-instances are also required for a proper training of data-intensive models. 
We also provide a web-accessible tool  that can be used to predict the voltage of any electrode material within a minute. 


\section{Acknowledgements}
We acknowledge the Material Project database for the data used in this work.  R. P. J  thanks Dr. Kai Trepte for careful proof reading of the manuscript. R. P. J also thanks  Dr. Zhen Zhou and Dr. Xu Zhang for providing the DFT  intercalation voltage for the voltage profile diagram. This work was supported by the Office of Basic Energy Sciences, US Department of Energy, DE-SC0005027 and DE-SC0019432.

\section{Supporting Information Available}
Scatter plots showing the performance of several ML models on the holdout test sets and the Na-test set,  color map showing the parameter tuning in Li-only data, a table with the voltages predicted by DNN, SVM and KRR for Na-set along with the DFT values. A table with the ML predicted voltage of Na- and K-based electrodes along with the DFT values for Li-based electrodes.

\bibliography{ML_battery}

\end{document}